\documentclass[final,twocolumn,times]{elsarticle}

\usepackage{graphicx}

\usepackage{amssymb}

\journal{Physica B}

\begin{document}

\begin{frontmatter}



\title{Role of anion size, magnetic moment, and disorder on the properties of the organic conductor $\kappa$-(BETS)$\sb{2}$Ga$\sb{1-x}$Fe$\sb{x}$Cl$\sb{4-y}$Br$\sb{y}$}


\author{E. Steven$^1$, H.B. Cui$^1$, A. Kismarahardja$^1$, J.S. Brooks$^1$, D. Graf$^1$, and H. Kobayashi$^2$}

\address{$^1$Physics Dept. and National High Magnetic Field Laboratory, Florida State University, Tallahassee Florida 32310, USA}
\address{$^2$Department of Chemistry, College of Humanities and
Sciences, Nihon University, Sakurajosui 3-25-40, Setagaya-Ku,
Tokyo 156-8550, Japan}
\begin{abstract}
	Shubnikov-de Haas and angular dependent magnetoresistance oscillations have been used to explore the role of anion size, magnetic moment, and disorder in the organic conductors $\kappa$-(BETS)$\sb{2}$GaBr$\sb{4}$ and $\kappa$-(BETS)$\sb{2}$FeCl$\sb{2}$Br$\sb{2}$ in the isomorphic class $\kappa$-(BETS)$\sb{2}$Ga$\sb{1-x}$Fe$\sb{x}$Cl$\sb{4-y}$Br$\sb{y}$. The results, combined with previous work, show correlations between the anion composition (Ga$\sb{1-x}$Fe$\sb{x}$Cl$\sb{4-y}$Br$\sb{y}$) and the superconducting transition temperature, effective mass, Fermi surface topology, and the mean free path.
\end{abstract}


\end{frontmatter}


\section{Introduction}
\label{Introduction}
Bis(ethylenedithio)tetraselenafulvalene (the organic donor ``BETS") is an important example where there is a substantial interaction between the conducting $\pi$-electrons in the donor and the localized d-electrons in the anion. Specifically, this has been dramatically demonstrated in $\lambda$-(BETS)$\sb{2}$FeCl$\sb{4}$ and $\kappa$-(BETS)$\sb{2}$FeBr$\sb{4}$ where field induced superconductivity emerges from an antiferromagnetic insulating\cite{Uji01} and a low field superconducting\cite{Konoike04} state respectively. By the substitution of Ga for Fe, and Br for Cl, the properties of these compounds can be tuned in terms of effective d-electron moment and anion size. In both cases, the substitutions (Ga for Fe and/or Br for Cl) leads to a reduction in the $\pi$-d electron interaction\cite{Uji03, Uji04}.

\begin{figure}[h]
\includegraphics[scale=.50,angle=0]{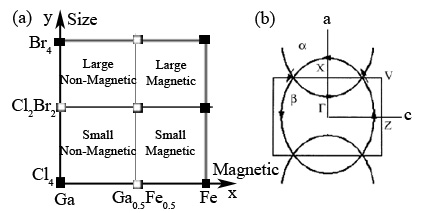}
\caption{(a) Alloying diagram of $\kappa$-(BETS)$\sb{2}$Ga$\sb{1-x}$Fe$\sb{x}$Cl$\sb{4-y}$Br$\sb{y}$:  black - measured, white- not yet synthesized alloy combinations; this work: GaBr$\sb{4}$ and FeCl$\sb{2}$Br$\sb{2}$. (b) Fermi surface  of $\kappa$-(BETS)$\sb{2}$FeCl$\sb{4}$\cite{Kobayashi96}. Lens orbit - $\alpha$, large orbit- $\beta$}
\label{fig:figure11}
\end{figure}

In the present work, we have investigated the alloy compound $\kappa$-(BETS)$\sb{2}$Ga$\sb{1-x}$Fe$\sb{x}$Cl$\sb{4-y}$Br$\sb{y}$ for two specific cases, $\kappa$-(BETS)$\sb{2}$GaBr$\sb{4}$ and $\kappa$-(BETS)$\sb{2}$FeCl$\sb{2}$Br$\sb{2}$ via the Shubnikov-de Haas (SDH) effect. From the SdH effect which probes the properties of charge carrier orbits on extremal areas of the quasi-two dimensional (Q1D) Fermi surface, information about the size of the Fermi surface orbits, the carrier effective mass, and the impurity scattering can be derived via the Lifshitz-Kosevich reduction factor relationships with magnetic field, field direction, and temperature\cite{Shoenberg84}. The Fermi surface of the $\kappa$-(BETS)$\sb{2}$GaBr$\sb{4}$ compound is shown in Fig.\ref{fig:figure11}, where two orbits, $\alpha$ and $\beta$, are present. The extremal area of the $\beta$-orbit in momentum space is equivalent to the area of the first Brillouin zone. The $\alpha$-orbit arises from the broken symmetry of the unit cell which leads to gap in the overlapping  $\beta$-orbit sections and hence the  $\beta$-orbit involves magnetic breakdown across the zone boundaries.

\section{Experimental Results}
\label{Experimental Results}

\begin{figure}[h]
\includegraphics[scale=.55,angle=0]{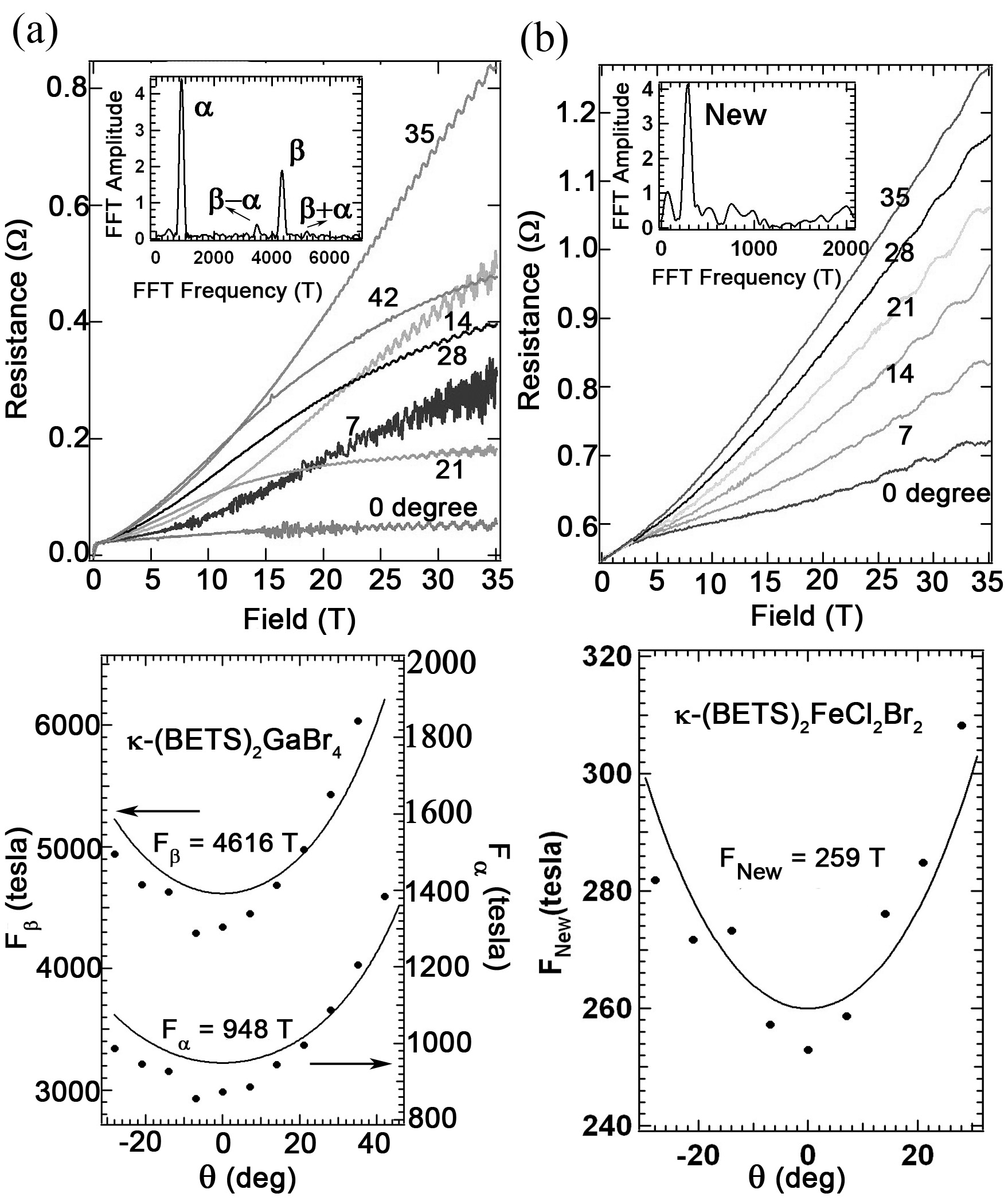}
\caption{Magnetoresistance (top) of $\kappa$-(BETS)$\sb{2}$GaBr$\sb{4}$ (a) and FeCl$\sb{2}$Br$\sb{2}$ (b) for different field directions. Inset: FFT showing the orbit frequencies. Angular dependence of the orbit frequencies compared with 1/cos$\theta$ dependence (bottom). }
\label{fig:figure22}
\end{figure}

\begin{figure}[h]
\linespread{1}
\par
\includegraphics[scale=.55,angle=0]{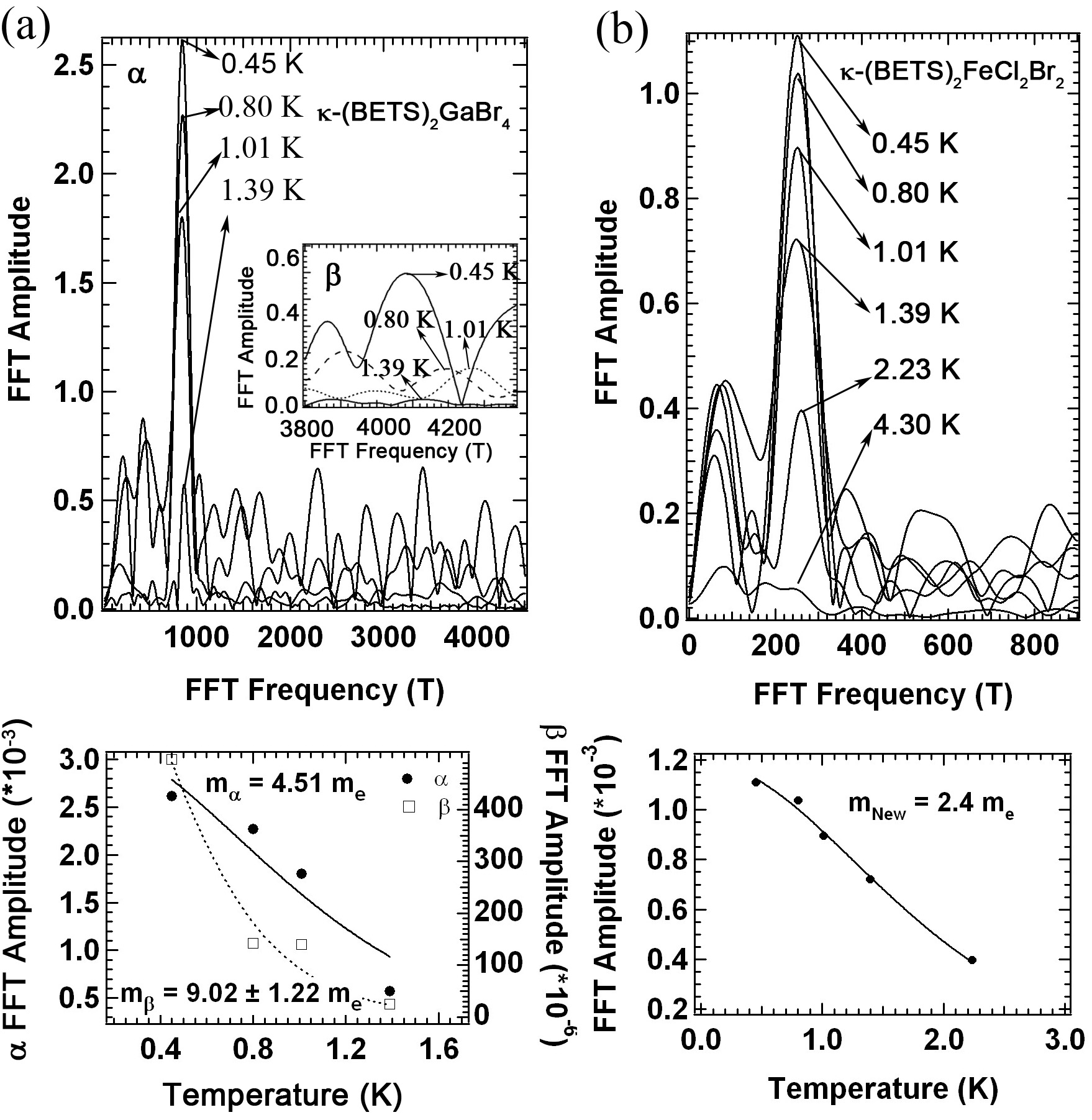}
\par
\caption{$\kappa$-(BETS)$\sb{2}$GaBr$\sb{4}$ (a) and $\kappa$-(BETS)$\sb{2}$FeCl$\sb{2}$Br$\sb{2}$ (b) FFT amplitude vs frequency at various temperatures for $\alpha$ orbit(top) and Lifshitz-Kosevich fit of FFT amplitude vs temperature data (bottom). Inset shows the FFT amplitude vs frequency at various temperatures for the $\beta$ orbit. }
\label{fig:figure33}
\end{figure}

Samples were grown by standard electrochemical methods. Electrical transport was by 4-terminal ac measurements with a current of order 10 $\mu$A along the b-axis of the single crystal samples where the b-axis is perpedicular to the a-c conducting plane. The samples were mounted on a rotatable probe using a standard helium cryostat and a 35 T resistive magnet.

The angular dependent magnetoresistance $\kappa$-(BETS)$\sb{2}$GaBr$\sb{4}$ and $\kappa$-(BETS)$\sb{2}$FeCl$\sb{2}$Br$\sb{2}$ is shown in Fig.\ref{fig:figure22} where $\theta$ is the angle between the magnetic field and the b-axis. The $\alpha$ and $\beta$ oscillations at 948 T and 4616 T respectively were observed for GaBr$\sb{4}$, but for FeCl$\sb{2}$Br$\sb{2}$ only a new oscillation frequency was observed at 259 T whose origin is not yet known. The oscillation frequencies all followed a quasi-two dimensional F$_0$/cos{$\theta$ relationship as shown in Fig.\ref{fig:figure22}.

	Fig.\ref{fig:figure33} shows the temperature dependence of the Fourier amplitudes of the oscillatory magnetoresistance as a function of magnetic field B, of both $\kappa$-(BETS)$\sb{2}$GaBr$\sb{4}$ and $\kappa$-(BETS)$\sb{2}$FeCl$\sb{2}$Br$\sb{2}$ for B//b where b is perpendicular to a and c axis shown in Fig.\ref{fig:figure22}. The cyclotron effective masses  were extracted  using the Lifshitz-Kosevich formulation as shown in Fig.\ref{fig:figure33}. For GaBr$\sb{4}$, the effective mass of the $\alpha$ orbit was  4.5 m$\sb{e}$. For the $\beta$ orbit, effective mass was estimated with a large error due to noise in the data to be 9 $\pm$ 1.22 m$\sb{e}$. For FeCl$\sb{2}$Br$\sb{2}$, the effective mass was 2.4 m$\sb{e}$ for the new orbit.

Fig.\ref{fig:figure44} shows the Shubnikov-de-Haas oscillations of both $\kappa$-(BETS)$\sb{2}$GaBr$\sb{4}$ and $\kappa$-(BETS)$\sb{2}$FeCl$\sb{2}$Br$\sb{2}$ at 0.45 K for  B//b. The fast Fourier transform (FFT) was applied to the oscillations with respect to the inverse field over a set of field intervals to obtain the Fourier amplitude vs. 1/B, which was then fit to the Lifshitz-Kosevich formula to obtain the Dingle temperatures (T$\sb{D}$) of the observed orbits. T$\sb{D}$'s for GaBr$\sb{4}$ were  0.55 K and 1.29 K for the  $\alpha$ and $\beta$ orbits respectively, while T$\sb{D}$ for  FeBr$\sb{2}$Cl$\sb{2}$ was 3.49 K.

Angular dependent magnetoresistance oscillation (AMRO) data for $\kappa$-(BETS)$\sb{2}$GaBr$\sb{4}$ at 10 and  18 T  (also at 20 and 35 T from Fig.\ref{fig:figure22}) is shown in Fig.\ref{fig:figure55}. The Fermi momentum k$\sb{F}$ for the $\beta$ orbit derived from the conventional  Tan($\theta$)  = n$\pi$ /b'k$\sb{F}$ relationship (where b' $\sim$ 36.635 / 2 $\AA$ \cite{Kobayashi93}) was $\sim$ 2.90 x 10$\sp{9}$ m$\sp{-1}$. The difference between the k$\sb{F}$ value extracted from AMRO and SdH data is due to our assumption of having a perfect circle Fermi surface when extracting k$\sb{F}$ from the SDH data.

\begin{figure}[h]
\linespread{1}
\par
\includegraphics[scale=.55,angle=0]{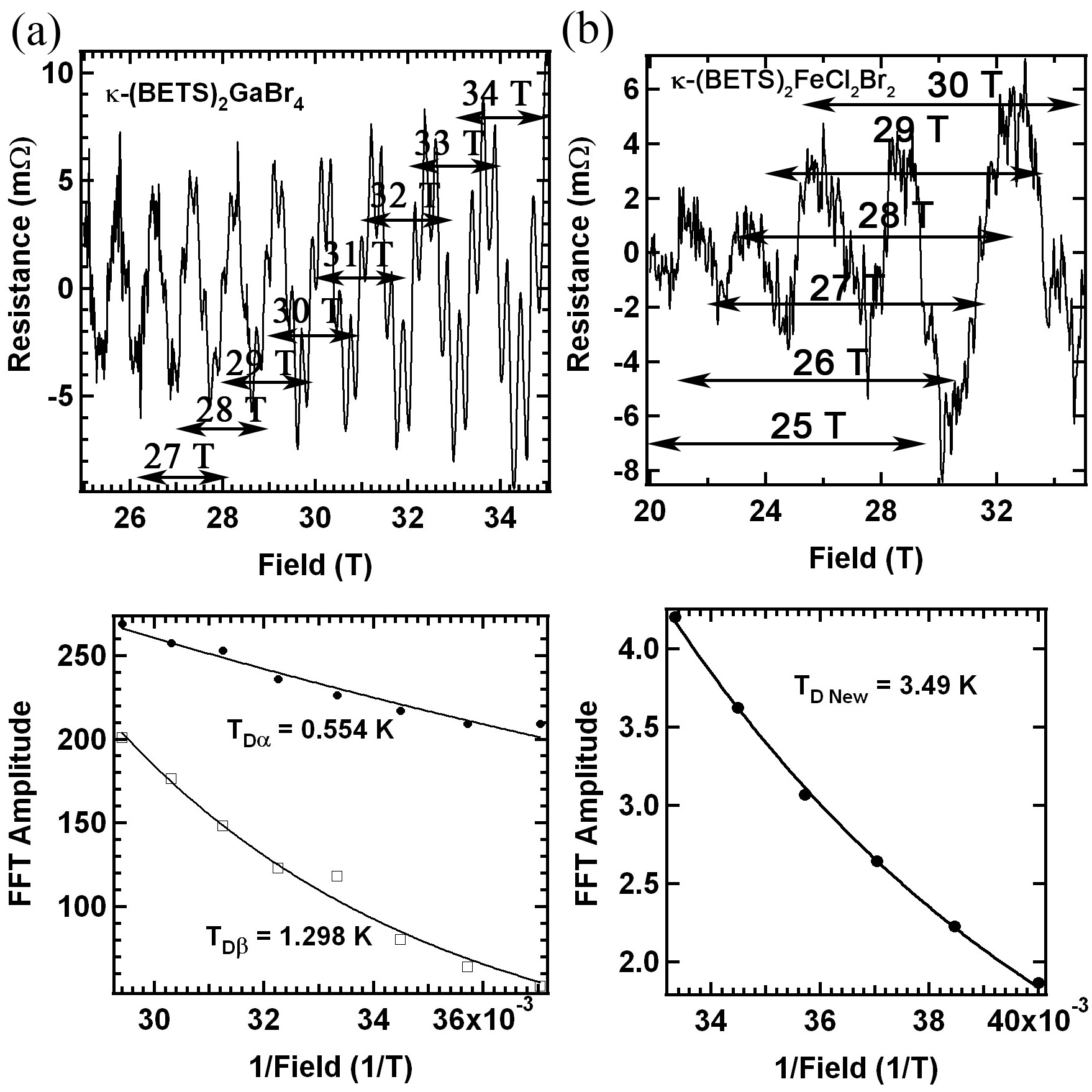}
\par
\caption{$\kappa$-(BETS)$\sb{2}$GaBr$\sb{4}$ (a) and $\kappa$-(BETS)$\sb{2}$FeCl$\sb{2}$Br$\sb{2}$ (b) magnetoresistance at B//b direction and 0.45 K (top)
and exponential fitting of FFT amplitude vs inverse field (bottom). Arrow shows the range at which FFT is performed at various field. }
\label{fig:figure44}
\end{figure}

\begin{figure}[p]
\includegraphics[scale=.85,angle=0]{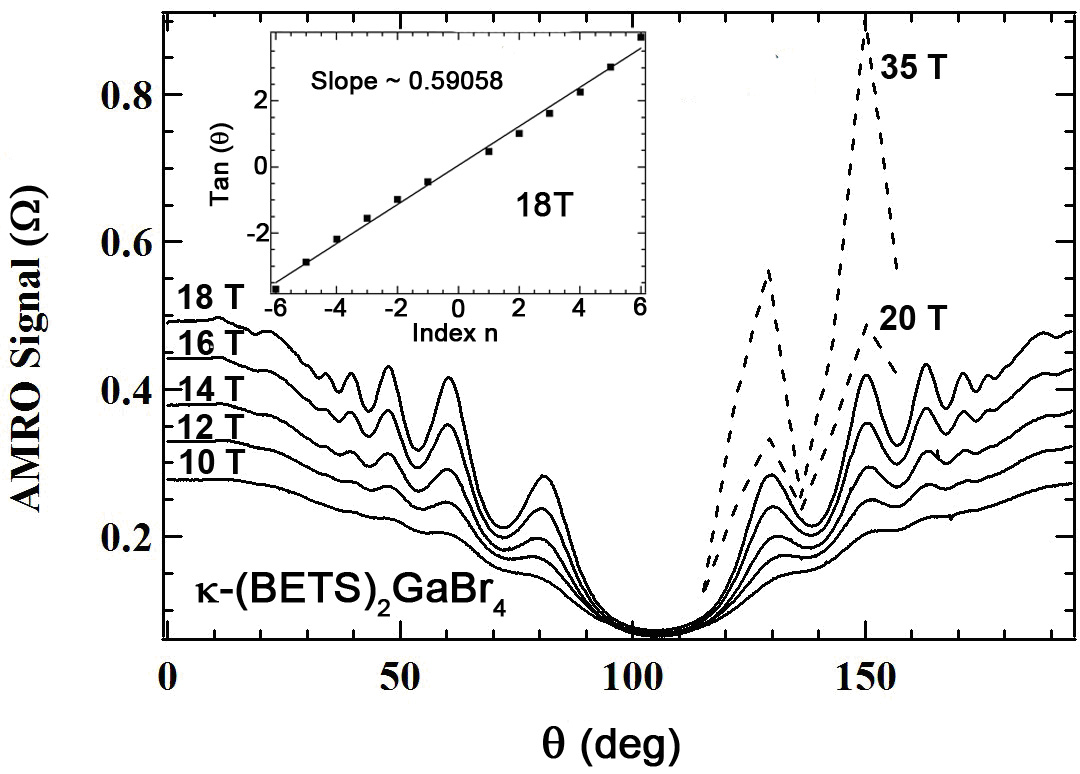}
\caption{ AMRO data of $\kappa$-(BETS)$\sb{2}$GaBr$\sb{4}$. Data taken at 10, 12, 14, 16 and 18 T, dotted line: AMRO extracted from Fig.\ref{fig:figure22} at 20 and 35 T. (Inset) Tan($\theta$) vs. index n with linear fitting using Tan($\theta$) $=$ n$\pi$ / b'k$\sb{F}$; b' is $\sim$ 36.635$\div$2 $\AA$ \cite{Kobayashi93}. k$\sb{F}$ for the $\beta$ orbit was extracted to be 2.90 * 10$\sp{9}$ /m.}
\label{fig:figure55}
\end{figure}

\begin{figure}[h]
\centering
\linespread{1}
\par
\includegraphics[scale=.45,angle=0]{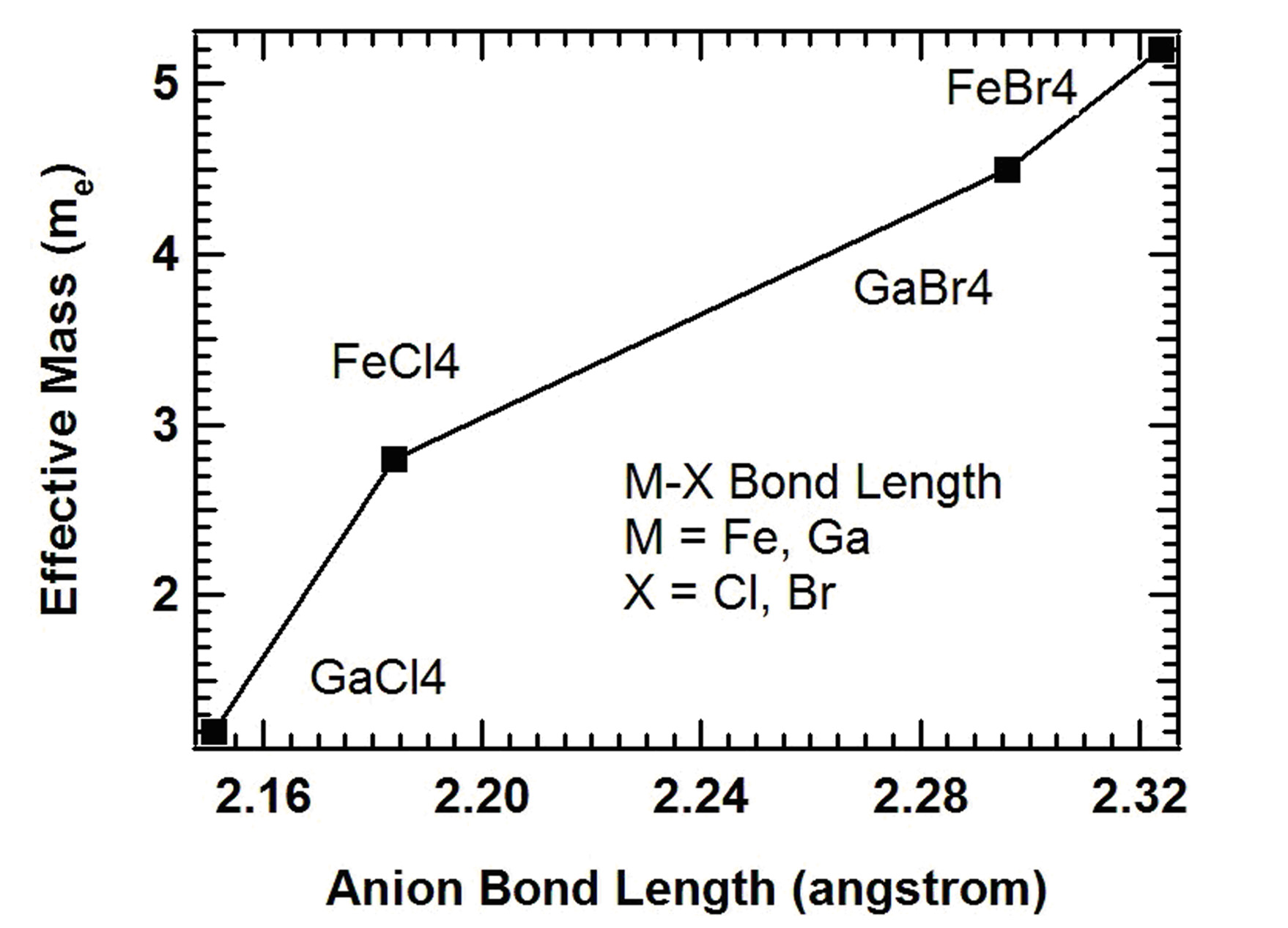}
\par
\caption{ Effective mass vs M-X anion bond length (M = Fe, Ga; X = Cl, Br). }
\label{fig:figure66}
\end{figure}

\section{Discussion}
\label{Discussion}
A comparison of our results with previous work\cite{Uji01, Harrison98, Tajima97} is shown in Table 1.  Several observations are of note. First, the effective mass appears to depend on the anion size (i.e. Cl vs. Br). Second, the effective mass also depends on the magnetic ion, rising about a factor of two between non-magnetic GaCl$\sb{4}$ and  magnetic FeCl$\sb{4}$, and about 20\% between GaBr$\sb{4}$ and FeBr$\sb{4}$. Third, a new oscillation frequency is observed in $\kappa$-(BETS)$\sb{2}$FeCl$\sb{2}$Br$\sb{2}$.

Harrison et al. \cite{Harrison98} considered the mass enhancement effect in terms a of spin fluctuation effect which allows the scaling of the electron-paramagnon interaction. Further evidence for  the interplay of conduction electron and magnetic moments in FeCl$\sb{4}$ is the negative magnetoresistance behavior in FeCl$\sb4$ \cite{Harrison98}, in contrast to the positive magnetoresistance   in GaCl$\sb4$ \cite{Tajima97},  suggesting that scattering is suppressed as the magnetic moments are aligned with field \cite{Harrison98}.  Furthermore, by comparing the positive magnetoresistance of GaBr$\sb4$ and the negative magnetoresistance of FeBr$\sb4$ \cite{Uji01}, it is likely that interaction between conduction electron and magnetic moments persists in FeBr$\sb4$  sample. The $\pi$-d interaction has been shown to be larger in purely Br samples as compared to pure Cl samples by Fujiwara et al \cite{Fujiwara02}. In their systematic study of FeBr$\sb{x}$Cl$\sb{4-x}$, they observed a pronounced maximum resistivity associated with strong $\pi$-d interaction for x = 4 and when x reaches $\sim$ 1.69 the maximum resistivity was no longer observable. In addition, we note that we have observed a possible correlation between the anion M-X bond length (M = Fe, Ga; X = Cl, Br) with the effective cyclotron mass as shown in Fig.\ref{fig:figure66}.

\begin{table*}[t]
\centering
\caption{Table 1. Summary of Fermi surface orbital frequencies and  corresponding effective cyclotron masses, respective anion bond length \cite{Kobayashi93}, unit cell volume \cite{Kobayashi93}, and superconducting transition temperatures.}
\begin{tabular*}{16.2cm}[t]{ p{2.6cm} p{2.2cm} p{2.2cm} p{2.7cm} p{2.2cm} p{2.0cm} }
  \hline\hline
   Orbits & GaCl$\sb{4}$\cite{Tajima96} & FeCl$\sb{4}$\cite{Harrison98} & GaBr$\sb{4}$ [This Work] & FeBr$\sb{4}$\cite{Uji01} & FeCl$\sb{2}$Br$\sb{2}$ [This Work] \\
  \hline
  $\alpha$ & 816 T, 1.2 m$\sb{e}$ & 864 T, 2.8 m$\sb{e}$ & 948 T, 4.5 m$\sb{e}$ & 850 T, 5.2 m$\sb{e}$ & None \\
  $\beta$ & 4350 T, 2.4 m$\sb{e}$ & 4296 T, 5.3 m$\sb{e}$ & 4616 T, 9$\pm$1.22 m$\sb{e}$ & 4280 T, 7.9 m$\sb{e}$ & None \\
  $\gamma$ & None & None & None & 103 T & None \\
  New & None & None & None & None & 259 T, 2.4 m$\sb{e}$ \\
  \hline
  Magnetic? & No & Yes & No & Yes & Yes \\
  \hline
  M-X bond length (M = Fe, Ga; X = Cl, Br) & 2.151 $\AA$ & 2.184 $\AA$ & 2.296 $\AA$ & 2.324 $\AA$ & - \\
  \hline
  Unit Cell Volume & 3544 $\AA^3$ & 3536 $\AA^3$ & 3663 $\AA^3$ & 3642 $\AA^3$ & 3613 $\AA^3$ \cite{Fujiwara02} \\
  \hline
  T$\sb{c}$ &  & 0.17 K\cite{Otsuka01} & 1 K\cite{Kobayashi02} & 1.1 K\cite{Ojima99} &  \\
  \hline\hline
\end{tabular*}
\label{table:table11}
\end{table*}

\begin{table*}[t]
\centering
\caption{Summary of Fermi surface and carrier lifetime parameters from this work.}
\begin{tabular*}{16.2cm}[t]{lccccccc}
  \hline\hline
     & Disorder & Area (10$\sp{18}$/m$\sp{2}$) & k$\sb{F}$ (10$\sp{9}$/m) & v$\sb{F}$ (10$\sp{4}$m/s) & T$\sb{D}$ (K) & $\tau$ (10$\sp{-12}$s) & $\ell$ (nm) \\
   \hline
   GaBr$\sb{4}$ ($\alpha$) & Low & 9.060 & 1.698 & 4.357 & 0.554 & 2.194 & 96 \\
   GaBr$\sb{4}$ ($\beta$) & Low & 44.08 & 3.746 & 4.805 & 1.298 & 0.937 & 45 \\
   FeCl$\sb{2}$Br$\sb{2}$ & Medium & 2.482 & 0.888 & 4.280 & 3.49 & 0.348 & 15 \\
   \hline\hline
\end{tabular*}
\label{table:table22}
\end{table*}

Correlation between the cyclotron effective mass and the superconducting temperature T$\sb{c}$ is further supported by an additional information from our GaBr$\sb{4}$ data which suggests that the cyclotron effective mass is proportional to T$\sb{c}$ as shown in Table.\ref{table:table11}. This correlation may be explained by the direct proportionality of both the effective mass $(N(0) = m^*/\pi\hbar^2)$ \cite{Smrcka95} and inverse exponential proportionality of the T$\sb{c}$ $(k_BT_c \sim \hbar\omega_De^{1/N(0)V})$ with the density of states at the Fermi level N(0).
 The origin of the new frequency in $\kappa$-(BETS)$\sb{2}$FeCl$\sb{2}$Br$\sb{2}$ is not yet known. The calculated mean-free-path as shown in Table.\ref{table:table22} shows that the FeCl$\sb{2}$Br$\sb{2}$ sample has a much higher disorder relative to the GaBr$\sb{4}$ sample. The calculated Fermi velocity, v$\sb{F}$, is almost the same for FeCl$\sb{2}$Br$\sb{2}$ and GaBr$\sb{4}$.

\section{Conclusions}
\label{Conclusions}
In general, substituting Cl with Br increases the effective cyclotron mass of the carriers. Simple arguments where the increase in lattice reduces the density of states at the Fermi level can account in part for this. Furthermore, similar effective mass enhancement by alloying the system with magnetic ion (Fe) in purely Cl sample has also been observed in purely Br system. It is likely that this effect is also due to the magnetic interaction in the system. A linear correlation between the anion bond length and effective cyclotron mass has been observed but more data is needed to confirm this point. The higher disorder system has lower mean-free-path due mostly to the decrease in the scattering time constant. The $\alpha$ and $\beta$ orbits has not been observed for FeCl$\sb{2}$Br$\sb{2}$ system but instead, a very small orbit of unknown origin has been observed. In addition, no negative magnetoresistance is observed, and the AMRO effect is also not evident. Clearly, the FeCl$\sb{2}$Br$\sb{2}$ compound differs in a fundamental way from the other compounds studied to date in the $\kappa$-(BETS)$\sb{2}$Ga$\sb{1-x}$Fe$\sb{x}$Cl$\sb{4-y}$Br$\sb{y}$ family.

\section{Acknowledgements}
\label{Acknowledgements}

This work is supported by NSF-DMR 06-02859. The
National High Magnetic Field Laboratory is supported by NSF
DMR-0654118, by the State of Florida, and the DOE.


\end{document}